Mars' formation can constrain the primordial orbits of the gas giants

Jason Man Yin Woo[1,2], Joachim Stadel[2], Simon Grimm[3], Ramon Brasser[4]

Abstract

Recent high precision meteoritic data infers that Mars finished its accretion rapidly within 10 Myr of the beginning of the Solar system and had an accretion zone that did not entirely overlap with the Earth's. Here we present a detailed study of the accretion zone of planetary embryos from high resolution simulations of planetesimals in a disc. We found that all simulations with Jupiter and Saturn on their current eccentric orbits (EJS) result in a similar accretion zone between fast-forming Mars and Earth region embryos. Assuming more circular orbits for Jupiter and Saturn (CJS), on the other hand, has a significantly higher chance of forming Mars with an accretion zone not entirely dominated by Earth and Venus region embryos, however CJS in general forms Mars slower than in EJS. By further quantifying the degree of overlap between accretion zones of embryos in different regions with the average overlap coefficient (OVL), we found that the OVL of CJS shows a better match with the OVL from a chondritic isotopic mixing model of Earth and Mars, which indicates that the giant planets are likely to have resided on more circular orbits than today during gas disc dissipation, matching their suggested pre-instability orbits. More samples, including those from Mercury and Venus, could potentially confirm this hypothesis.

[1] Corresponding author: jasonwoohkhk@gmail.com

[2] Institute for Computational Science, University of Zürich, Winterthurerstrasse 190, 8057 Zürich, Switzerland

[3] Center for Space and Habitability, University of Bern, Gesellschaftsstrasse 6, 3012 Bern, Switzerland

[4] Earth Life Science Institute, Tokyo Institute of Technology, Meguro-ku, Tokyo 152-8550, Japan

1. Introduction

Recently improved isotopic measurements of martian meteorites, mainly from the most abundant shergottite-nakhlite-chassignite (SNC) meteorites, allows us to constrain the bulk composition of Mars. Early oxygen isotopic mixing models suggested that Mars is composed of more than 80% ordinary chondrite (OC) like material (Lodders and Fegley 1997). Recent improvements of laboratory apparatus allows high precision isotopic measurements on lower-concentration nucleosynthetic isotopes, such as $^{48}$Ca, $^{54}$Cr, $^{50}$Ti, etc. Chondritic isotope mixing models involving these nucleosynthetic isotopes provide a more complete comparison between Earth's and Mars' bulk composition. Assuming chondritic material as the building blocks of the terrestrial planets, recent models suggest that Mars accreted about half of its mass from OC-like material, and another half from enstatite chondrite (EC) like material (Sanloup et al. 1999; Tang & Dauphas 2014; Brasser et al. 2018), whereas Earth accreted more than 70% EC-like material and the remaining 30% mostly from OC-like material (e.g Javoy, 2010; Dauphas 2017). To account for the isotopic differences between Earth and Mars, Mars is suggested to have formed in a different region from that of the Earth, potentially even in the asteroid belt (Brasser et al. 2017).

Isotopic measurements of martian meteorites not only constrain Mars' bulk composition, but also its formation timescale. Among all the radioactive decay systems, the $^{182}$Hf-$^{182}$W system is the most widely adopted for dating the core-formation time of a body because of its ideally short half-life (8.9 Myr) and the nature of the parent lithophile isotope decaying to its daughter siderophile isotope. By applying the Hf-W chronology, Dauphas & Pourmond (2011) showed that Mars accreted very rapidly; approximately 50% of its mass after 2 Myr and finished most of its accretion between 5 to 10 Myr after the formation of the Solar System. This result is supported by another radioactive decay system $^{60}$Fe-$^{60}$Ni (Tang & Dauphas 2014). On the other hand, Earth's accretion took much longer. Yin et al. (2002) and Kleine et al. (2009) showed that Earth's accretion and core formation lasts about ~30 Myr to > 100 Myr based on the Hf-W chronology.

Recent studies applied the above constraints on Earth's and Mars' composition to test different existing dynamical models with *N*–body simulations (Woo et al. 2018; Mah & Brasser 2021). Combining the martian Hf-W chronology with *N*–body simulations is, however, more challenging, since this requires much higher resolution simulations to model Mars' formation from a disc of planetesimals. Thanks to the increasingly powerful computational performance of CPUs and GPUs, simulations of terrestrial planet formation starting from the pre-runaway stage has become possible (e.g. Walsh & Levison 2019; Clement et al. 2020). In our previous work (Woo et al. 2021), we performed simulations starting with planetesimals in the classical model (i.e. the solid surface density follows the minimum mass Solar Nebular (MMSN); Hayashi 1981) and the depleted disc model (i.e. the solid surface density is reduced beyond Mars' orbit relative to the MMSN; Izidoro et al. 2014) to study the growth rate of Mars. We discovered the following three conditions favour a faster growth for Mars: (1) the giant planets are placed on their current orbits (eccentric Jupiter-Saturn, EJS), instead of on more circular orbits (CJS) as suggested by the Nice model (Tsiganis et al. 2005); (2) the gas disc decay timescale is ≤ 1 Myr and (3) Mars mainly formed from larger planetesimals (because fewer collisions are required to form Mars-sized embryos when the planetesimal diameter is larger). Due to a much higher degree of mixing present for embryos forming in the EJS model (condition 1), Earth and Mars then tend to have a rather similar composition. This would contradict the observed isotopic differences between Earth and Mars.

The purpose of this letter is to follow up on the study of Woo et al. (2021), analysing the accretion zones of embryos case by case. We perform a more quantitative comparison between the accretion zones of the fast-forming Mars region embryos and those in the Earth region, then compare results to the isotopic mixing model.

It is currently still unclear whether the giant planets possessed eccentric or circular orbits during the gas disc phase. Although the traditional Nice model assumes circular orbits (CJS) for the giant planets (Tsiganis et al. 2005), some hydrodynamical simulations result in eccentric Jupiter and Saturn if they migrate in the gas disc and become trapped in mean-motion resonances (Pierens et al. 2014). Using the final orbital configuration of the gas giants from their hydrodynamical simulations, Pierens et al. (2014) claim that they successfully reproduce the outer solar system architecture. Also, if the giant planet's cores formed via pebble accretion, involving mutual scattering between planetesimals (Levison et al. 2015), these cores and their fully formed giant planets may not be in very circular orbits (e < 0.01). Furthermore, if gas giants formed from gravitational instability of gas fragments in the protosolar disc, which could be an alternative path for giant planet formation (e.g. Boss, 1997; Vorobyov and Elbakyan, 2018), gas giants' initial eccentricities are not well determined. Hence, we study both EJS and CJS scenarios.

## 2. Method

Here we briefly summarize the simulation set up from Woo et al. (2021). In Appendix A we outline the initial conditions used for simulations of both the classical (Chambers 2001) as well as the depleted disc model (Izidoro et al. 2014). The GPU *N*-body code *GENGA* (Grimm & Stadel 2014) allows direct simulations with several tens of thousands of fully self-gravitating objects by increasing performance through specialised and self-tuning GPU kernels. We analyse results from 28 runs of the classical model and 13 runs of the depleted disc model.

The gas disc model is based on Morishima et al. (2010) with a decay timescale $\tau_\text{decay}$ = 1 or 2 Myr (See Equation (1) of Woo et al. (2021)). The effect of decaying gas disc gravitational potential, aerodynamic gas drag (Adachi et al., 1976) and Type-I migration (Tanaka & Ward 2004) are included. All simulations were run with a timestep of 5 days for 10 Myr. During close encounters and collisions the effective timestep is much shorter, as a direct Bulirsch-Stoer integration is used in such cases. The fully grown giant planets are placed on either EJS or CJS orbits. For the depleted disc model we were only able to perform simulations for the EJS case due to limitations of computational resources.

## 3. Results

We found that most Mars embryos do not grow fast enough in our simulations (Woo et al. 2021) when compared to the timescale suggested by Dauphas & Pourmand (2011). However, this timescale from Dauphas and Pourmand (2011) is model dependent, as they extrapolates their measurement in the context of Mars accretion following the simplistic oligarchic growth curve of Chambers (2006). On the other hand, the formation of embryos in our simulations is directly modeled, as the embryos build up naturally from smaller planetesimals. Given this model dependence in the suggested timescale of Dauphas and Pourmand (2011), we slightly relax the constraints by defining an embryo which reaches 0.8 times the current Mars mass, in the current Mars region (1.25 AU < *a* < 1.75 AU) at 10 Myr, as a "fast-forming Mars embryo".

### 3.1. Statistics of fast-forming Mars embryos

Among all the EJS classical simulations with $\tau_{\text{decay}}$ = 2 Myr, > 70 % (5 of 7) formed at least one fast-forming Mars embryo whereas this number decreases to ~8% (1 of 13) for the depleted disc model (See Appendix B for the detailed statistics). This difference is due to the lower initial solid surface density beyond the current orbits of Mars and the asteroid belt in the depleted disc model. The growth rate (d$M$/d$t$) of an embryo is proportional to the local solid surface density (e.g. Kobayashi & Dauphas 2013). Hence, in general, Mars region embryos grow more slowly in the depleted disc than in the classical model.

The probability of yielding fast-forming Mars embryos can be increased if we decrease $\tau_{\text{decay}}$ from 2 Myr to 1 Myr. We found that about ~86 % (6 of 7) of the EJS classical simulations with $\tau_{\text{decay}}$ = 1 Myr have at least one fast-forming Mars embryo, which is slightly higher than for the $\tau_{\text{decay}}$ = 2 Myr case. It is because the $\nu_5$ secular resonance with Jupiter sweeps through the asteroid belt during gas dissipation earlier (dubbed "sweeping secular resonance"; e.g. Bromley & Kenyon 2017; Nagasawa et al. 2000; Hoffmann et al. 2017). This process injects the asteroid belt material into the terrestrial planet region. As a result, the solid surface density in the inner Solar System is increased and thus the formation of embryos in the terrestrial planet region is sped up at an earlier stage in a shorter lived gas disc.

Comparing EJS and CJS for the classical simulations, CJS has a lower chance (~57 %) (8 of 14) of producing a fast-forming Mars embryo. It is because Jupiter's initial eccentricity is an order of magnitude lower in CJS than in EJS and hence the $\nu_5$ secular resonance is much weaker in CJS. The implantation of asteroid belt material into the terrestrial planet region by sweeping secular resonance is negligible in CJS, leading to a generally slower formation rate of embryos in the Mars region in CJS.

To conclude, the classical model has a much higher probability (~80% for EJS) of yielding fast-forming Mars embryos than the depleted disc model (<10% for EJS), because of the higher solid surface density beyond the Mars region. For the classical cases, EJS results in a higher probability of rapidly forming Mars region embryos than for the CJS simulations. Nevertheless, we still have ~57% of CJS simulations yielding at least one fast-forming Mars embryo, which is not low. However, the final system architecture may not match the sizes and separations of the terrestrial planets. We will study the system architecture in a future paper.

## 3.2. Accretion zone analysis

The sweeping secular resonance, however, also induces severe mixing within the terrestrial planet region, causing the accretion zones of embryos to be similar to each other. In the following, we present the accretion zone of the fast-forming Mars embryos in comparison with the accretion zone of embryos in other regions in each simulation (Appendix C gives the precise definition of the accretion zone which we used). We define the Mercury, Venus, Earth and Mars regions as 0.27 AU < $a$ < 0.5 AU, 0.5 AU < a < 0.85 AU, 0.85 AU < $a$ < 1.25 AU and 1.25 AU < $a$ < 1.75 AU.

Figure 1 depicts the accretion zone of embryos for each EJS simulation with $\tau_{\text{decay}}$ = 2 Myr. Five runs are from the classical model ((a) to (e)) and 1 run is from the depleted disc model (f). For Mars region embryos, we plot only the data for fast-forming embryos since they are the ones that are more likely to match the Hf-W chronological data from the martian meteorites (Dauphas & Pourmand 2011; Tang & Dauphas 2014). It is obvious that in every run the embryo accretion zones show substantial overlap. We are most interested in the comparison between Earth region embryos and fast-forming Mars embryos since the existing meteoritic data indicate that their accretion zones should be different (Dauphas et al. 2017; Brasser et al. 2017). This is not the case in Figure 1: we observe that the fast-forming Mars embryos all have accretion zones covered entirely by those of the Earth region embryos. This is also true for the comparison between Venus region embryos and Mars region embryos. Since the subsequent giant impact phase (Agnor et al. 1999; Chambers, 2001) involves scattering between embryos and the viscous stirring caused by the embryos, which dynamically heats up the disc even further, we expect the final planetary system would retain this highly mixed state until the assembly of the terrestrial planets is completed.

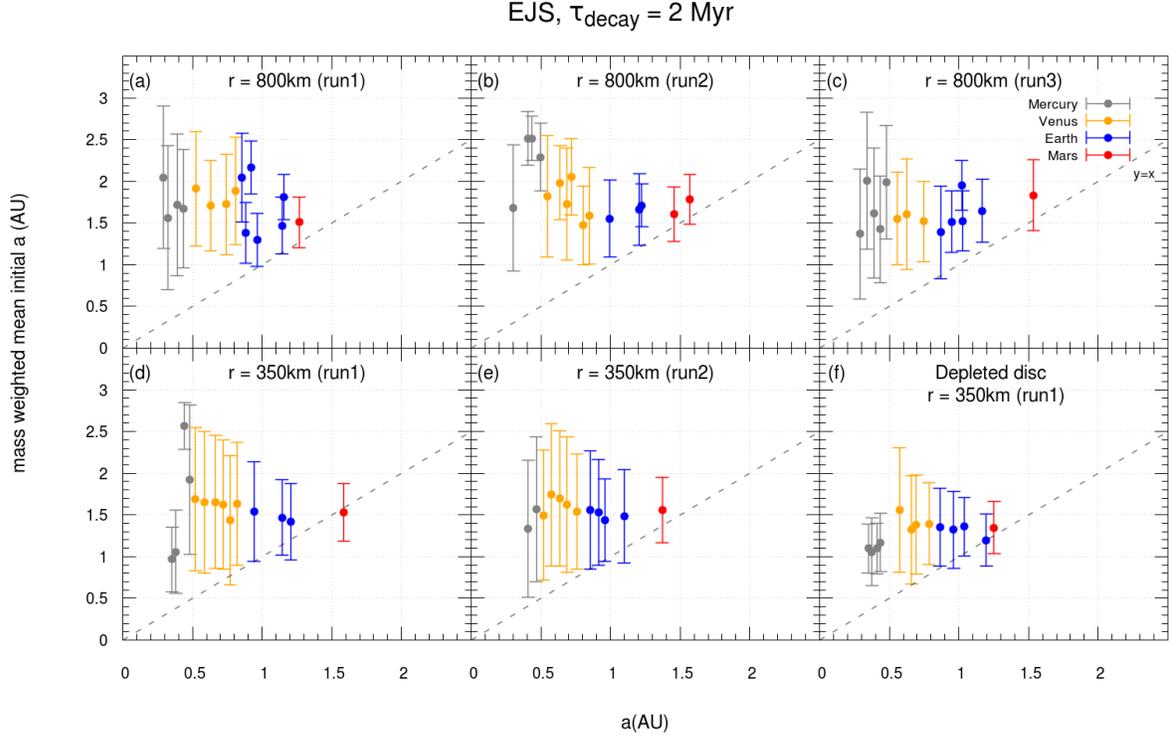

Figure 1 - Mass weighted mean initial semi-major axis of embryos in the current region of Mercury (grey), Venus (green), Earth (blue) and Mars (red) after 10 Myr for each simulation. The error bars represent the zone over which material is mainly accreted to form a particular embryo (see Appendix C). We plot only the embryos that reached 0.8 $M_{Mars}$ at 10 Myr for those in the current Mars region. The gas disc decay timescale $\tau_{decay}$ is 2 Myr and the gas giants are on their current eccentric orbits (EJS). The initial radii of the planetesimals ($r$) are either 800 km or 350 km. The simulations (a) to (e) begin with a solid surface density following the minimum mass Solar nebula (MMSN, Hayashi 1981). The lower right case (f) shows the depleted disc model. In this last case we chose a depletion location ($a_{dep}$) at 1.5 AU, beyond which the disc is depleted by 75 % ($\beta$ = 75 %) with respect to 1.5 times of the MMSN (See Appendix B for the detailed description of the initial conditions). The dashed lines are the linear correlation line y=x.

Figure 2 shows the accretion zones of embryos in EJS simulations with a shorter disc lifetime ($\tau_{decay}$ = 1 Myr). It is clear that decreasing $\tau_{decay}$ to 1 Myr does not have a significant effect on the results. Similar to Figure 1, in every run the accretion zone of the fast-forming Mars embryo is entirely covered by Earth and Venus region embryos. This shows that the sweeping secular resonances, and not Type-I migration (Tanaka et al 2002; Tanaka & Ward 2004), is the dominant effect for the extensive mixing in EJS simulations, since Type-I migration is less effective in a gas disc with a shorter decay timescale. On the other hand, the mixing effect of sweeping secular resonances remains similar between long lived and short lived discs (comparing Figures 1 and 2).

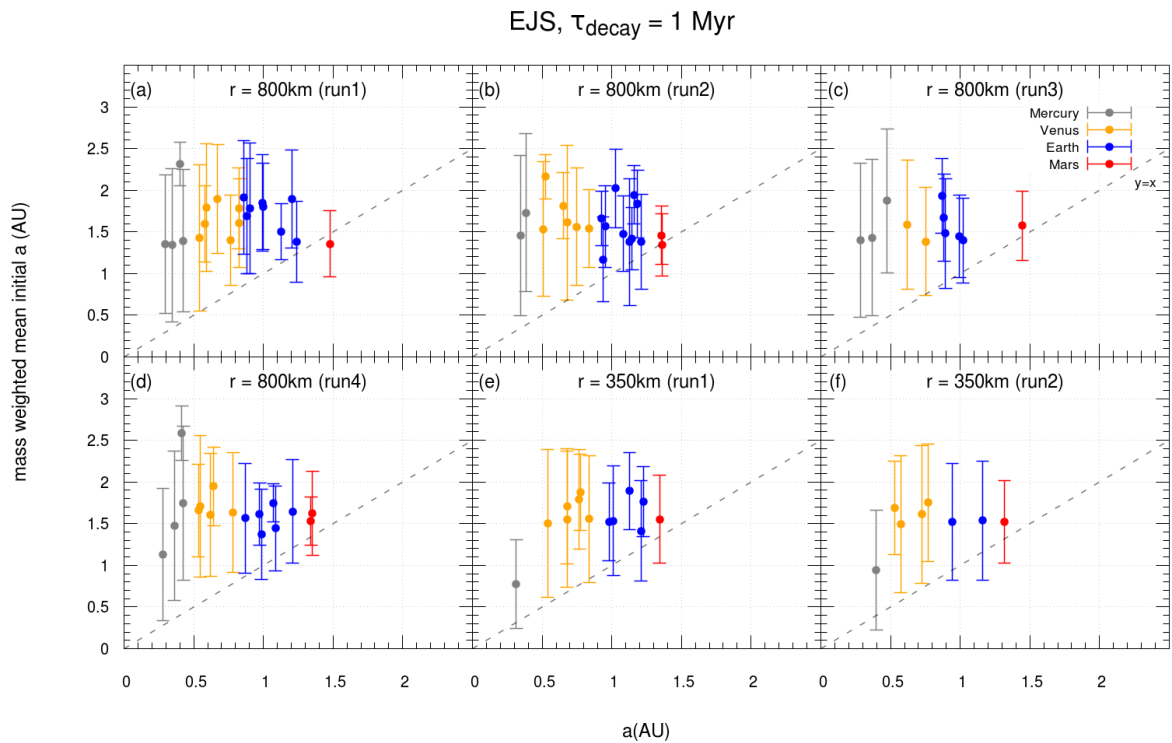

Figure 2 – Same as Figure 1, but for $\tau_{decay}$ = 1 Myr and all the cases begin with a solid surface density following the MMSN (the classical model).

Placing the giant planets on more circular orbits, has a greater impact on the outcome of accretion zones. Figure 3 shows the accretion zones of embryos in every CJS simulation having at least one fast-forming Mars embryo. Similar to the combined results presented in Woo et al. (2021), all runs show mostly local accretion. Also, unlike the EJS case nearly all embryos show a narrow (<1 AU) accretion zone and a much lower degree of mixing. Comparing the fast-forming Mars and Earth region embryos in each run, we found that in 5 out of 8 runs the fast-forming Mars embryo has accreted from the most distant region. This shows that unlike EJS, CJS is much more likely to generate a fast-forming Mars with an accretion zone not entirely covered by the Earth region embryos (see Section 3.3).

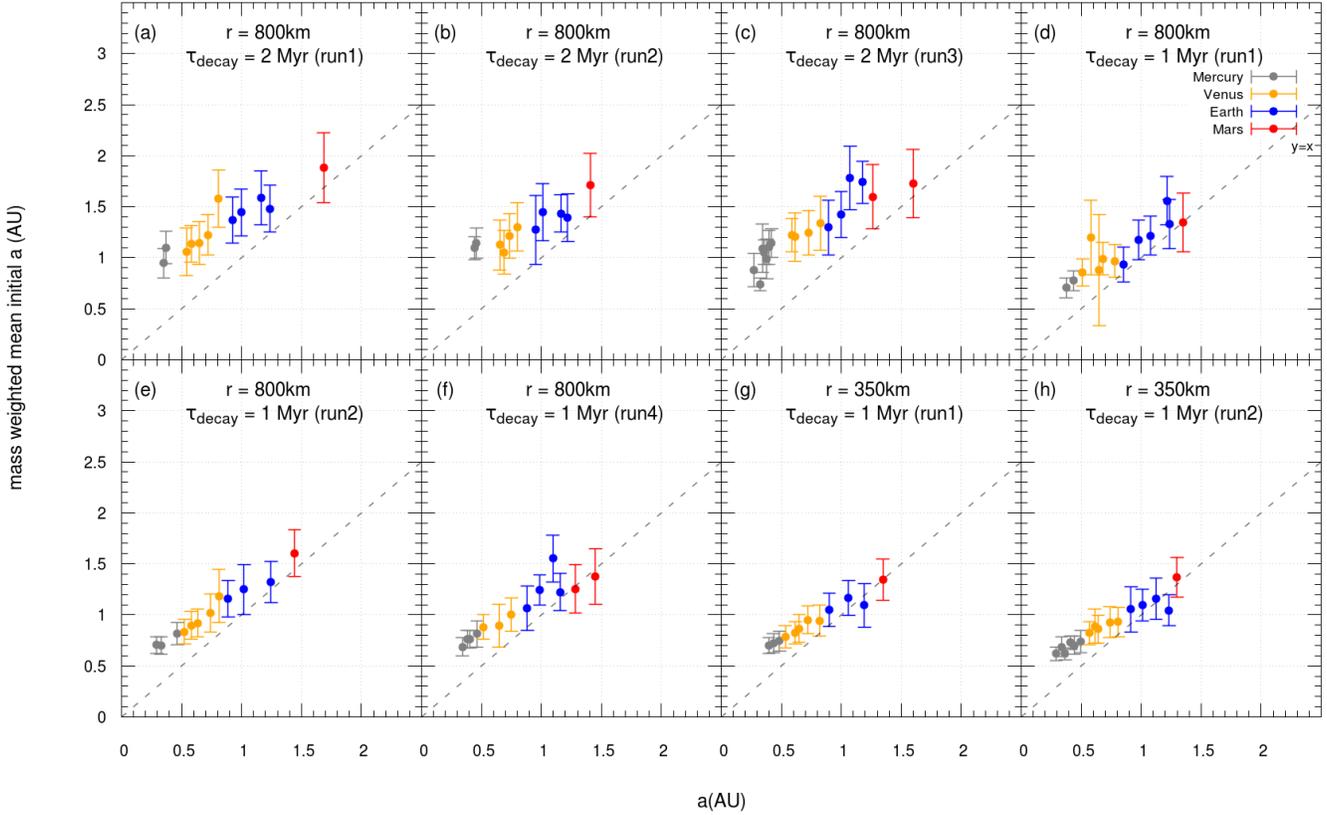

Figure 3 - Same as Figure 2, but for a more circular Jupiter and Saturn (CJS) and $\tau_{decay}$ = 1 or 2 Myr. All such CJS simulations, regardless of resolution and of gas disc lifetime, maintain a low degree of mixing during embryo formation.

To summarise, the mixing effect in the EJS simulations is much stronger than in CJS, resulting in a highly overlapping accretion zone between fast-forming Mars embryos and Earth and Venus region embryos in EJS simulations; whereas the accretion zone of embryos is narrower and less overlapping in the CJS simulations. Thus, CJS simulations have a higher success rate in forming Earth and Mars region embryos with differing final composition.

### 3.3. The average overlap coefficient (OVL)

In order to quantify the degree of overlap between the accretion zones of embryos in different regions, we compute the average OVL (Appendix D) for three sets of simulations: (1) EJS, $\tau_{decay}$ = 2 Myr (Figure 1); (2) EJS, $\tau_{decay}$ = 1 Myr (Figure 2) and (3) CJS (including both $\tau_{decay}$ = 1 and 2 Myr, Figure 3). Table 1 shows the average OVL for each embryo pair in different simulation sets. The larger the OVL the higher degree of overlap between the accretion zones of two regions' embryos. All the embryo pairs have the highest OVL for EJS, $\tau_{decay}$ = 2 Myr. Decreasing $\tau_{decay}$ to 1 Myr tends to decrease the OVL for the Mercury pairs, since more material is retained in the innermost region of the disc with $\tau_{decay}$ = 1 Myr. Embryos in the Mercury region accrete mostly from this innermost material when $\tau_{decay}$ = 1 Myr, whereas this same material is lost to the Sun due to longer-lasting Type-I migration when $\tau_{decay}$ = 2 Myr. This gas disc decay time difference leads to higher variations of the accretion zone of Mercury region embryos with embryos from other regions in EJS $\tau_{decay}$ = 1 Myr than $\tau_{decay}$ = 2 Myr. The same does not apply to other embryo pairs as there is almost no difference in their OVL between $\tau_{decay}$ = 1 Myr and 2 Myr. As expected, the CJS simulations have the lowest OVL for most pairs (except Mercury-Venus) because of the lower degree of mixing in the disc.

While the results for Mercury region embryos are interesting, we focus on the Earth-Mars pairs since there is existing data from samples that allows direct comparison. Based on the chondritic isotopic mixing model of Dauphas (2017) and with the expansion to a greater variety of chondrites, Mah & Brasser (2021) showed that the OVL for Earth-Mars is 0.58 ± 0.08. Our results imply that CJS with a lower OVL = 0.56 ± 0.12 is more probable in reproducing the limited overlapping between Earth and a fast-forming Mars embryo composition implied from meteoritic data. Still we cannot rule out the results from EJS if we include the 1σ uncertainties.

Table 1 – The average overlap coefficient (OVL) of each embryo pair in each simulation set.

|  | Mercury-Venus | Mercury-Earth | Mercury-Mars | Venus-Earth | Venus-Mars | **Earth-Mars** |
|---|---|---|---|---|---|---|
| EJS, $\tau_{decay}$ = 2 Myr | 0.69 ± 0.03 | 0.46 ± 0.08 | 0.32 ± 0.12 | 0.68 ± 0.07 | 0.49 ± 0.11 | **0.70 ± 0.08** |
| EJS, $\tau_{decay}$ = 1 Myr | 0.50 ± 0.11 | 0.30 ± 0.09 | 0.16 ± 0.05 | 0.68 ± 0.07 | 0.46 ± 0.06 | **0.69 ± 0.06** |
| CJS | 0.53 ± 0.11 | 0.24 ± 0.12 | 0.10 ± 0.07 | 0.55 ± 0.09 | 0.28 ± 0.11 | **0.56 ± 0.12** |
| Chondritic mixing model | - | - | - | - | - | **0.58 ± 0.08** |

Note - The uncertainties are 1 standard deviation. The chondritic mixing model for Earth-Mars are from Mah & Brasser (2021). We highlighted the Earth-Mars pair since it is the only pair that can be compared directly to the chondritic isotopic mixing model.

## 4. Conclusions

We performed over 40 state-of-the-art *N*-body simulations to study the formation of embryos directly from a disc of only planetesimals. We focus on whether embryos in the Mars region can form rapidly (reaching ~0.8 $M_{Mars}$ within 10 Myr) while simultaneously having an accretion zone different from Earth region embryos. We found that the CJS simulations produce fast-forming Mars embryos with less Earth-overlapping accretion zones (Figure 3) and results in an overlap coefficient (OVL) closely matching the chondritic mixing model's result (Table 1). This indicates that even though the EJS model has higher probability in forming Mars fast (~80 %, whereas in CJS ~57 %) and clearing the mass in the Mars region and the asteroid belt due to sweeping secular resonances, its severe mixing effect could violate the isotopic differences between Earth and Mars (e.g. Sanloup et al. 1999) and the inferred heliocentric isotopic gradient in the inner Solar System (Yamakawa et al. 2010; Mezger et al. 2020). The CJS case, on the other hand, matches predictions for the giant planets' orbital state during gas disc dispersal (e.g. Tsiganis et al. 2005) and also the isotopic data from various meteorites including the martian meteorites that we showed in this study. Hence, we postulate that the giant planets were likely to possess more circular orbits than today during gas disc dispersal. As suggested in our previous paper (Woo et al. 2021), due to the differences of the accretion zones of embryos in the Mercury and Venus region between EJS and CJS simulations, samples from Mercury and Venus could settle this debate.

The remaining challenge is to solve the problems of CJS, which produces a Mars and an asteroid belt that are both too massive (Chambers 2001; O'Brien et al 2006; Raymond et al. 2009), as well as producing too many low mass planets in the terrestrial planet region (Clement et al. 2020). One of the proposed solutions is to have the giant planets undergo an early instability phase in < 100 Myr after gas disc dispersal (e.g. Clement et al. 2018; Ribeiro et al. 2020). Radiogenic reset ages from the asteroid belt support an instability before 80 Myr (Mojzsis et al., 2019). Giant planet migration and subsequent instability excites the planetesimals and subsequently clears them around the Mars and asteroid belt region (e.g. Clement et al. 2018, 2019). Clement et al. (2020) suggested this instability may also excite the terrestrial planets region and help overcome the problem of having too few giant impacts during the embryo collision phase, which results in forming too many planets in the terrestrial planet region in their simulations. However, the giant planet's instability may also induce mixing within the disc, which could possibly erase the composition gradient developed in our CJS simulations (Figure 3). Future simulations need to be carried out in order to verify this claim.

There are factors that we did not consider in our simulations. Fragmentation, which has been suggested to speed up the initial growth of the embryos during the gas disc phase (Kobayashi & Dauphas 2013), should be implemented in our future simulation. Inducing a slight excitement of the dynamical state to the planetesimals by gas disc turbulence (Laughlin et al. 2004; Ogihara et al. 2007) may also help in speeding up the growth of embryos in the Mars region, so as to more closely match with the timescale inferred by Hf-W dating. A comprehensive set of future high-resolution simulations including various dynamical effects could help solve the puzzles of the growth rate, composition and architecture of the terrestrial planets and constrain the early evolution of the giant planets.

## Acknowledgment

This work has been carried out within the framework of the National Center of Competence in Research PlanetS, supported by the Swiss National Science Foundation (SNSF). The authors acknowledge the financial support of the SNSF. The authors acknowledge the computational support from Service and Support for Science IT ($S^3$IT) of University of Zurich and the Swiss National Supercomputing Centre (CSCS).

## Appendix A
Initial conditions of the simulations

We run simulations for both the classical model (e.g. Chambers, 2001; O'Brien et al., 2006; Raymond et al., 2009) and the depleted disc model (Izidoro et al., 2014; Mah and Brasser, 2021). In the classical model, the solid surface density profile $\Sigma_s = 7$ gcm$^{-2}$ $(a/\ 1\ \text{AU})^{-3/2}$, which follows the minimum mass Solar nebula (MMSN, Hayashi, 1981). In the depleted disc model, the solid surface density profile is given as the following:

$$\Sigma(a) = \begin{cases} \chi \Sigma_s (a/\ 1\ \text{AU})^{-3/2}; & a \leq a_{\text{dep}}, \\ (1-\beta) \chi \Sigma_s (a/\ 1\ \text{AU})^{-3/2}; & a > a_{\text{dep}}, \end{cases} \quad (A1)$$

where $a_{\text{dep}} = 1$, 1.25 or 1.5 AU is the depletion location, $\beta = 50\%$ or 75% is the depletion factor and $\chi = 1$, 1.5 or 2 is the scaling factor of MMSN. The parameters are chosen from simulations in Mah and Brasser (2021) that achieved the highest probability of forming Mars with the correct mass. Table A1 shows the initial conditions of our simulations for this study.

Table A1 - The initial conditions of our classical and depleted disc simulations for this study. We study the growth of embryos from a disc of planetesimals with different simulation set-ups, including planetesimals' radii ($r$), total number of initial planetesimals ($N$), total mass of the initial disc, decay timescale of the gas disc ($\tau_{\text{decay}}$) and orbits of the giant planets - eccentric Jupiter-Saturn (EJS) and circular Jupiter-Saturn (CJS).

|  | $a_{\text{dep}}$, $\beta$, $\chi$ (Equation A1) | Planetesimals' radii ($r$) | No. of initial planetesimals ($N$) | Mass of initial disc (in $M_{\text{Earth}}$) | $\tau_{decay}$ | Giant planets' orbit | Number of simulations |
|---|---|---|---|---|---|---|---|
| Depleted | 1.5 AU, 50%, 1 | 350 km | 28168 | 2.5 | 2 Myr | EJS | 2 |
|  | 1.5 AU, 75%, 1.5 | 350 km | 35307 | 3.2 | 2 Myr | EJS | 2 |
|  | 1.25 AU, | 350 km | 41235 | 3.7 | 2 Myr | EJS | 2 |
| disc | 75%, 2 | | | | | | |
| | 1 AU, 75%, 2 | 350 km | 34763 | 3.1 | 2 Myr | EJS | 2 |
| | 1.5 AU, 50%, 1 | 800 km | 2337 | 2.5 | 2 Myr | EJS | 5 |
| Classical | NIL | 350 km | 37437 | 3.4 | 2 Myr | EJS | 2 |
| | | 350 km | 37437 | 3.4 | 2 Myr | CJS | 2 |
| | | 350 km | 37437 | 3.4 | 1 Myr | EJS | 2 |
| | | 350 km | 37437 | 3.4 | 1 Myr | CJS | 2 |
| | | 800 km | 3114 | 3.4 | 1 Myr | EJS | 5 |
| | | 800 km | 3114 | 3.4 | 2 Myr | EJS | 5 |
| | | 800 km | 3114 | 3.4 | 2 Myr | CJS | 5 |
| | | 800 km | 3114 | 3.4 | 1 Myr | CJS | 5 |

Appendix B

Statistics of fast-forming Mars embryo

We present the statistics of forming embryos in the Mars region quick enough in different sets of simulations. According to the Hf-W chronology of Dauphas and Pourmand (2011), Mars is likely to finish accreting most of its mass within 10 Myr of the beginning of the Solar System. We thus define an embryo that reaches 0.8 $M_{\text{Mars}}$ in the current Mars region (1.25 AU < $a$ < 1.75 AU) at 10 Myr as the "fast-forming Mars embryo". Table A2 shows the percentage of simulations that yield at least one fast-forming Mars embryo. We have discussed the detailed statistics in Section 3.1.

Table A2 - Statistic of yielding fast-forming Mars embryos in different sets of depleted disc and classical simulations. We combine the high resolution ($r$ = 350 km) and low resolution data ($r$ = 800 km) for each simulation set. See Table A1 for the detailed simulation sets we performed.

|  | Simulation sets | Total no. of simulations | Total no. of simulation with at least one fast-forming Mars embryo | Percentage of simulation with at least one fast-forming Mars embryo |
|---|---|---|---|---|
| Depleted disc | EJS, $\tau_{decay}$ = 2 Myr | 13 | 1 | 7.7 % |
| Classical | EJS, $\tau_{decay}$ = 2 Myr | 7 | 5 | 71.4 % |
| Classical | EJS, $\tau_{decay}$ = 1 Myr | 7 | 6 | 85.7 % |
| Classical | CJS, $\tau_{decay}$ = 1 and 2 Myr | 14 | 8 | 57.1 % |

Appendix C

Definition of the accretion zone

We define the mass weighted mean initial semi-major axis, $a_{weight}$ as the mean of the accretion zone of each embryo. It is written as:

$$a_{\text{weight}} = \frac{\sum_i^{N_{\text{acc}}} m_i a_i}{\sum_i^{N_{\text{acc}}} m_i} \quad (A2)$$

(Brasser et al., 2017; Fischer et al., 2018; Kaib and Cowan, 2015; Woo et al., 2018), where $m_i$ and $a_i$ are the initial mass and semi-major axis of the accreted planetesimals and $N_{acc}$ is the number of planetesimals accreted by the embryo. Its mass weighted standard deviation is written as

$$\sigma_{\text{weight}} = \sqrt{\frac{\sum_i^{N_{\text{acc}}} m_i (a_i - a_{\text{weight}})^2}{((N_{\text{acc}}-1)/N_{\text{acc}}) \sum_i^{N_{\text{acc}}} m_i}} \quad (A3)$$

(Brasser et al., 2017; Kaib and Cowan, 2015; Mah and Brasser, 2021; Woo et al., 2018). The accretion zone for each embryo in Figure 1, Figure 2 and Figure 3 is represented by $a_{weight} \pm \sigma_{weight}$.

Appendix D

Calculation of the average overlap coefficient (OVL)

In order to calculate the average OVL of each set of simulations in Table 3, we proceed as follows:

1) Obtain the initial semi-major axes of the accreted planetesimals for the embryos in different regions.
2) Combine the initial semi-major axes of the accreted planetesimals for embryos in the same region into one file.
3) Bin the data with an interval of 0.1 AU and calculate the mass fraction of planetesimals originating from each bin for each embryo region. The mass fraction of planetesimals is relative to the total mass of embryos in that region.
4) Comparing the mass fraction of planetesimals originated from each bin between embryos from two different regions, take the number with a smaller mass fraction and subsequently summarize this mass fraction for each bin. For example, if we are comparing the OVL between Venus region embryos and Earth region embryos in a particular simulation, the OVL can be calculated as:

$$\text{OVL} = \sum_1^n \min[f_{\text{Venus}}, f_{\text{Earth}}], \quad (A4)$$

where $f_{\text{Venus}}$ and $f_{\text{Earth}}$ is the fraction of mass that originated from a particular bin (e.g. 0.5-0.6 AU, 0.6-0.7 AU, …, 2.9 to 3 AU) for embryos in Venus region and the Earth region, respectively, and $n$ is the total number of bins.
5) Repeat step (4) for different embryos pairs in different simulations.
6) Take the average and 1$\sigma$ standard deviation of the OVL for each pairs in different simulations for a simulation set (three sets: (1)EJS, $\tau_{\text{decay}}$ = 2 Myr; (2) EJS, $\tau_{\text{decay}}$ = 1 Myr and (3) CJS (including both $\tau_{\text{decay}}$ = 1 and 2 Myr)). The final results are presented in Table 1.

References

Adachi, I., Hayashi, C., Nakazawa, K., 1976, PThPh, 56, 1756

Agnor, C. B., Canup, R. M., & Levison, H. F. 1999, Icar, 142, 219

Boss, A. P., 1997, Sci, 276, 1836

Brasser, R., Mojzsis, S. J., Matsumura, S., & Ida, S. 2017, E&PSL, 468, 85

Brasser, R., Dauphas, N., & Mojzsis, S. J. 2018, GeoRL, 45, 5908

Bromley, B. C., Kenyon, S. J., 2017, AJ, 153, 216

Chambers, J. E. 2001, Icar, 152, 205

Chambers, J. E. 2006, Icar, 180, 496

Clement, M. S., Kaib, N. A., Raymond, S. N., Walsh, K. J., 2018. Icar, 311, 340

Clement, M. S., Raymond, S. N., Kaib, N. A., 2019, AJ, 157, 38

Clement, M. S., Kaib, N. A., Chambers, J. E., 2020, PSJ, 1, 18

Dauphas, N., Pourmand, A., 2011, Natur, 473, 489

Dauphas, N., 2017, Natur, 541, 521

Fischer, R. A., Nimmo, F., O'Brien, D. P., E&PSL, 482, 105

Grimm, S. L., Stadel, J. G., 2014, ApJ, 796, 23

Hayashi, C., 1981, PThPh 70, 35

Hoffmann, V., Grimm, S. L., Moore, B., Stadel, J., 2017, MNRAS, 465, 2170

Izidoro, A., Haghighipour, N., Winter, O. C., Tsuchida, M., 2014, ApJ, 782, 31

Javoy, M., Kaminski, E., Guyot, F. et al., 2010, E&PSL, 293, 259

Kaib, N. A., Cowan, N. B., 2015, Icar, 252, 161

Kleine, T., Touboul, M., Bourdon, B., et al., 2009, GECoA, 73, 5150

Kobayashi, H., Dauphas, N., 2013, Icar, 225, 122

Lodders, K., Fegley, B., 1997, Icar, 126, 373

Laughlin, G., Steinacker, A., Adams, F. C., 2004, ApJ, 608, 489

Levison, H.F., Kretke, K.A., Duncan, M.J., 2015, Natur, 524, 322

Mah, J., Brasser, R., 2021, Icar, 354, 114052

Mezger, K., Schönbächler, M., Bouvier, A., 2020, SSRv, 216, 27

Morishima, R., Stadel, J., Moore, B., 2010, Icar, 207, 517

Mojzsis, S.J., Brasser, R., Kelly, N.M., Abramov, O., Werner, S.C., 2019, ApJ, 881, 44

Nagasawa, M., Tanaka, H., Ida, S., 2000, AJ, 119, 1480

O'Brien, D.P., Morbidelli, A., Levison, H.F., 2006, Icar, 184, 39

Ogihara, M., Ida, S., Morbidelli, A., 2007, Icar, 188, 522

Pierens, A., Raymond, S.N., Nesvorny, D., Morbidelli, A., 2014, ApJ, 795, L11

Sanloup, C., Jambon, A., Gillet, P., 1999, PEPI, 112, 43

Raymond, S. N., O'Brien, D. P., Morbidelli, A., Kaib, N., 2009, Icar, 203, 644

Ribeiro, R. de S., Morbidelli, A., Raymond, S. N., et al. 2020, Icar, 339, 113605

Tang, H., Dauphas, N., 2014, E&PSL, 390, 264

Tanaka, H., Takeuchi, T., Ward, W. R., 2002, ApJ, 565, 1257

Tanaka, H., Ward, W. R., 2004, ApJ, 602, 388

Tsiganis, K., Gomes, R., Morbidelli, A., Levison, H. F., 2005, Natur, 435, 459

Vorobyov, E.I., Elbakyan, V.G., 2018, A&A, 618, A7

Walsh, K. J., Levison, H. F., 2019, Icar, 329, 88

Woo, J. M. Y., Brasser, R., Matsumura, S., Mojzsis, S. J., Ida, S., A&A, 2018, 617, A17

Woo, J. M. Y., Grimm S., Brasser R., Stadel J., 2021, Icar, 359, 114305

Yamakawa, A., Yamashita, K., Makishima, A., Nakamura, E., 2010, ApJ, 720, 150

Yin, Q., Jacobsen, S. B., Yamashita, K., et al. 2002, Natur, 418, 949